\documentclass[epl,twocolumn,floatfix,groupedaddress]{revtex4}

\usepackage{graphicx,amsfonts,amssymb,hyperref,nicefrac,lipsum, verbatim}

\usepackage[fleqn]{amsmath}
\usepackage{mathtools}

\usepackage[x11names]{xcolor}
\newif\ifhyper
\hypertrue
\ifhyper
\hypersetup{
   citecolor = {green!60!black},
   colorlinks = {true}, 
   urlcolor = {blue} 
}
\fi

\newcommand{\be}{\begin{equation}}
\newcommand{\ee}{\end{equation}}
\newcommand{\beqa}{\begin{eqnarray}}
\newcommand{\eeqa}{\end{eqnarray}}
\newcommand{\ket} [1] {\vert #1 \rangle}
\newcommand{\bra} [1] {\langle #1 \vert}
\newcommand{\braket}[2]{\langle #1 | #2 \rangle}

\def\bra#1{\langle#1\vert}
\def\ket#1{\vert#1\rangle}

\def\Longarrow{\protect\@lra}
\def\@lra{\relbar\joinrel\relbar\joinrel\relbar\joinrel%
          \relbar\joinrel\rightarrow}

\renewcommand{\vec}[1]{\bm{#1}}

\begin{document}

\title{Topological Magnetic Excitations}


\author{M.~Malki}
\affiliation{Lehrstuhl f\"ur Theoretische Physik 1, TU Dortmund, Germany}

\author{G.\ S.~Uhrig}
\affiliation{Lehrstuhl f\"ur Theoretische Physik 1, TU Dortmund, Germany}

\date{\rm\today}

\begin{abstract}
Topological properties play an increasingly important role in future research and technology. This also applies to the field of topological magnetic excitations which has recently become a very active and broad field. In this perspective article, we give an insight into the current theoretical and experimental investigations and try an outlook on future lines of research.
\end{abstract}

\maketitle

\section{Introduction}

Quantum magnetism continues to fascinate researchers in condensed matter physics. 
Insulating quantum magnets provide  clean 
solid state systems which are characterized by only a few coupling constants.
The magnetic degrees of freedom do not interact strongly with other degrees of freedom
which has enabled a multitude of fundamental and applied achievements. 
Applications range from the compass needle in the distant past to
 deflection magnets in beamlines and magnetic data storage 
indispensable in today's information technologies \cite{fert08}. 
The next major revolution in this respect could be not only to store the information magnetically, but also to transmit and to process it by magnetic excitations.

Usually, signal transmission and procession are accompanied by 
dissipative losses. 
Topological systems have the outstanding feature to be robust against many
perturbations, e.g., by suppressing backscattering and thereby reducing dissipation.
This can enhance the efficiency of devices decisively. Since the discovery of the quantum Hall effect
\cite{klitz80,storm83},  topological effects have appeared in condensed matter systems in various forms. Topological magnetic excitations are one important route of research 
in this area with promising potential for magnetic devices. Using neutral magnetic excitations instead
of charge excitations avoids long-range interactions with stray electric fields which act
as noise source. Thus, less dissipative devices based on topological magnetics can be hoped for.

In the light of the above, this perspective article gives a survey over the current research on
magnetic excitations with topological properties focussing on localized spins in insulators
or magnetic semiconductors. It is organized in three parts. First, the theoretical point of view is presented. Second, the experimental state-of-the-art is discussed. Finally, we conclude with 
an outlook. Of course, a complete account over all activities in this
booming research field is not possible.

\section{Theoretical Foundations}

The symmetries of a system determine the possibility of non-trivial topology. The lower the symmetry,
the more coupling terms are possible which may give the
eigenstates a twist such that topological characteristics arise. For magnetic insulators the
generic twisting couplings are the Dzyaloshinskii--Moriya (DM) interactions
\begin{align}
\label{eq:dm}
\mathcal{H}_{ij,\mathrm{DM}} = \vec{D}_{ij} \cdot (\vec{S}_i \times \vec{S}_j)
\end{align}
resulting from spin-orbit coupling on the electronic \smash{level \cite{shekh92,shekh93}.}
The above mentioned symmetry aspects are given by Moryia's rules \cite{moriy60b} which constraints the direction a DM vector $\vec{D}_{ij}$ can point. The selection rules and how
spin-orbit coupling leads to DM terms are shown in the Supplemental Sects.\ \ref{supp:one} and \ref{supp:two}.

An alternative coupling which can also lead to non-trivial topology is the
scalar spin chirality \cite{shind01}
\begin{align}
\label{eq:ssc}
\mathcal{H}_{ijk,\mathrm{SSC}} = \chi \, \vec{S}_{i} \cdot (\vec{S}_j \times \vec{S}_k) , 
\end{align}
involving three different spins and favoring their non-coplanar orientation. 
The cross product in both terms \eqref{eq:dm} and \eqref{eq:ssc}
implies the required magnetic anisotropy. 
The scalar spin chirality does not necessarily require spin-orbit coupling 
\cite{zhang17, zhang18}. 
Expressing the spin operators in terms of bosons allows one to address
the elementary excitations, the quasi-particles, as bosons \cite{xu20, kumar20}. 

\begin{figure}
\centering
\includegraphics[width=\columnwidth]{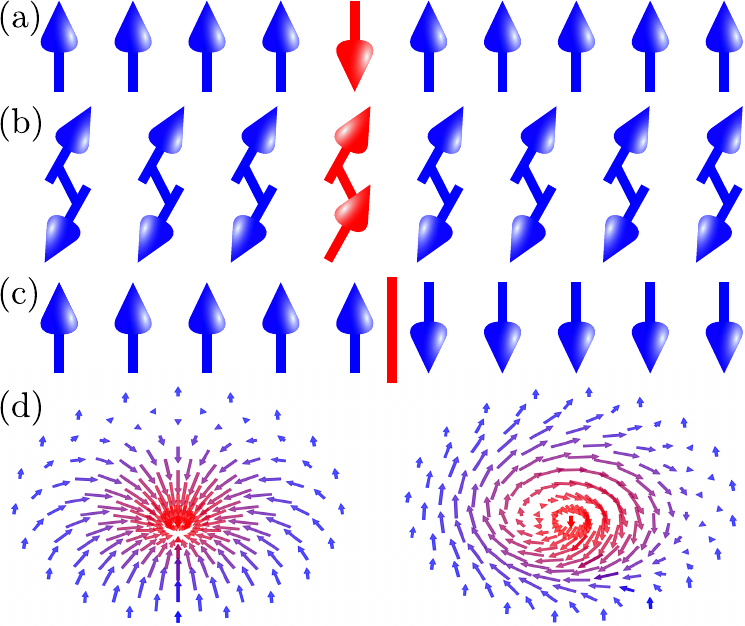} 
\caption{Sketch of a magnon (a), a triplon (b), a spinon (c) and a skyrmion excitation (d). 
The blue and red bonds in (b) correspond to a singlet and a triplet state, respectively. 
\smash{Panel (d)} shows a N{\'e}el (left) and a Bloch (right) skyrmion.}
\label{fig:exc}
\vspace{-0.3cm}
\end{figure}

On the level of a single particle, bosons and fermions behave in the same way so that features of
fermionic models carry over to bosonic and magnonic ones, see also Supplemental Sect.\ \ref{supp:three}.
Common representations for ordering quantum magnets
are the ones of Holstein and \smash{Primakoff  \cite{holst40},} of Dyson and Maleev 
\cite{dyson56a, malee57} and of \smash{Schwinger \cite{schwi66}} which
 describe quantized spin waves: magnons \cite{bloch30}. Disordered magnets
can be valence bond solids with triplon excitations \cite{knett00a,schmi03c,knett03a}
which can be captured by bond \smash{operators \cite{sachd90,norma11}.}
In one dimension (1D) and for strong frustration,  fractionalization of the 
magnetic excitations to spinons \cite{fadde81} can arise. In 1D they can be understood
as domain walls. They are intricate objects and 
do not allow for straightforward bosonic descriptions, see for instance Ref.\ \cite{hafez17a}. Additionally, long-range ordered magnets 
can host stable solitonic static
spin waves such as skyrmions. These four types of magnetic excitations form the 
body of this survey and are schematically shown in Fig. \ref{fig:exc}. 
The magnon excitation is the most common magnetic quasi-particle and 
therefore takes the largest part of 
the remainder. Note that the ground states are  mostly
topologically trivial, independent of the nature of the excitations.

The Berry phase is the basic concept in defining topological properties. 
In the general context of differential geometry, it is the change of a vector in
a fibre bundle upon parallel transport along a closed path on the manifold.
In quantum mechanics, it appears as 
the phase that an eigenstate of a parameter dependent 
Hamiltonian acquires when the parameters are changed along a closed path
\cite{berry84}, see Supplemental Sect. \ref{supp:four}. 
Depending on the relevant parameter space several invariants can be 
defined by Berry phases. The closed path around the Brillouine zone  in two dimensions (2D) defines the Chern number
\begin{align}
\label{eq:chern}
C_n := \frac{1}{2 \pi} \iint_{\mathrm{BZ}} F_{n, xy} (\vec{k}) \mathrm{d} k_x \mathrm{d} k_y ,
\end{align}
where $F_{n, xy} (\vec{k})$ is the Berry curvature of the band $n$ expressed by
\begin{align}
F_{n, xy} (\vec{k}) := 
\mathrm{i} \left(\braket{\partial_{k_x} n}{\partial_{k_y} n} - 
\braket{\partial_{k_y} n}{\partial_{k_x} n} \right).
\label{eq:berry-curva}
\end{align}
The wave vectors $k_x$ and $k_y$ parametrize the manifold,
here the torus of the 2D Brillouin zone..
The key feature of the Chern number is that it can only take
integer values so that it cannot change its value continuously. 
This implies that at boundaries between phases of different
Chern numbers the system changes non-perturbatively and
thus energy gaps must close. This leads to the famous bulk-boundary
correspondence implying the existence of in-gap states 
\cite{berne13} which are often localized boundary states
if protected by indirect energy gaps in the bulk \cite{malki19b}.
Another integer topological invariant derived from 
the Berry phase is the Zak phase in 1D systems
with inversion symmetry \cite{zak89}. In addition, winding numbers
characterize topological objects beyond Berry phases. They are
based on covering mappings of closed manifolds onto spheres of various 
dimensions \cite{fruch13, asbot16, chiu16}.

\subsection{Magnon excitations}

A fermionic lattice model with a non-trivial Chern number was first proposed by Haldane \cite{halda88b}
implying a quantized transversal conductance. Exploiting the analogy, Haldane and Arovas \cite{halda95} 
transferred the concept also to a quantum spin system proposing transversal spin current due to a finite 
Chern number in a 2D chiral disordered gapped quantum magnet. Then it took 15 years before 
conventional ordered magnets were considered for topological transport \smash{phenomena \cite{fujim09}.} 
The prediction of a thermal Hall effect \cite{katsu10a} induced by 
topological magnon or spinon bands has triggered studies on topological quantum magnets since 
the tranversal heat conductivity is accessible in experiments. 
Topological magnons were first proposed on the kagome lattice induced by
scalar spin chirality \cite{katsu10a, owerr17}. 
Magnon bands in an insulating kagome ferromagnet (FM) with DM interactions were calculated subsequently
\cite{mook14b, mook14, chisn15, sesha18}. An external magnetic field can be added 
to fully polarize the ground state, to break time-reversal symmetry or to open energy gaps between bands. 

Further FM lattices with DM interactions have been analyzed to show topological magnons with non-trivial Chern numbers such as the pyrochlore \cite{zhang13}, the Shastry--Sutherland \cite{malki19b}, the 
honeycomb \cite{owerr16, kim16, owerre16c}, the bilayer honeycomb \cite{owerr16b} and the Lieb lattice \cite{cao15}. 
Studies of periodic arrays of FM islands have shown that magnetic dipole-dipole interactions can also lead to
 topological magnons on square and honeycomb lattices \cite{shind13b, shind13, wang17}. 
The elementary excitations of antiferromagnetically (AFM) ordered models are magnons as well, but
magnons in AFMs differ from their FM counterparts in that they have a linear dispersion at low
wave number and that their number is not conserved. 
In second quantization, this also requires to pass from the standard scalar product
to an adapted symplectic product, also called `paraunitary', see e.g.\ \smash{Ref.\
\cite{malki19c}} and references therein.
Non-trivial Chern numbers of AFM magnon bands 
have been investigated in the honeycomb \cite{owerr16, owerr17c}, the pyrochlore \cite{laure17}, the 
\smash{kagome \cite{laure18, mook19},} the Shastry--Sutherland \cite{bhowm20} and the triangular \cite{kim19} lattice. 

The Kitaev model \cite{kitae06} is strongly frustrated due to the bond-directional couplings which 
favor exotic quantum states such as quantum spin liquids. The required anisotropy for non-trivial Chern numbers is intrinsically present if Kitaev interactions are combined
with FM or AFM couplings \cite{joshi18, mcclar18}.

The bulk-boundary correspondence \cite{berne13,mook14} predicts
chiral magnonic edge modes for non-trivial Chern \smash{numbers \cite{shind13}.} Here, the term `chiral' 
means that the excitations propagates only in one direction along one edge \cite{matsu11, mats11b}
like, e.g.,  skipping orbits in the quantum Hall effect.
This non-reciprocity is caused by the Berry curvature which acts on the excitations like a magnetic field
on charges.  A temperature gradient unbalances the occupation of magnons at two opposite edges implying
a thermally driven transversal spin currents, the so-called spin Nernst effect
\cite{mook14b, koval16, zyuzi16, owerr16b, owerr17, nakat17b, wang18b, lee18}.

The magnonic analogue of a fermionic Chern insulator is not the only kind of a magnetic topological phase. Magnons
can also characterize a topological semi-metal phases where the term
`metal' is to be understood only as analogy to fermionic systems. 
Topological semi-metals are divided in three categories:  Weyl, Dirac and nodal line semi-metals. 
These systems are characterized by finite quantized Berry phases around 
point degeneracies (monopoles) or line degeneracies (nodal lines)
in momentum space. Their boundary states are Fermi arcs for monopoles 
and drumhead surface states for nodal lines.
Magnonic Weyl systems have been proposed for 3D pyrochlore FMs and AFMs 
\cite{li16c, mook16, su17, jian18}, a stacked honeycomb FM \cite{su17b, zyuzi18}, and a stacked kagome AFM
\cite{owerr18b, owerr18c}. If time-reversal symmetry and inversion symmetry are preserved 
each energy on the Dirac cone is doubly degenerate. By analogy, the magnetic
counterpart of a Dirac semimetal has been predicted \cite{frans16, persh18, owerr2017d, boyko18}.
Topological nodal lines indicate the degeneracy of two bands along this line in momentum space induced
by topology, i.e., the Berry phase around such line does not vanish, but takes integer
values. Magnonic analogues have also been found \cite{mook17, li17, owerr17e, boyko18}.
 Lastly, a 3D  topological insulator with magnons has also been suggested \cite{li18}. 

Topological magnons in 1D can be characterized by the Zak phase, i.e., the 
Berry phase along the 1D Brillouin zone. So far, only a few magnetic chain systems are suggested 
with topologically non-trivial magnons \cite{qin17c, qin18b, pirmo18}.
The bulk-boundary correspondence predicts the existence of 
end states within the indirect gap between magnon  branches of non-trivial
Zak phase, cf.\ Ref.\ \cite{malki19c}.

In addition to static topological models, constant in time, 
first ideas on Floquet engineering \cite{eckar17} have been advocated. 
The magnetic dipole moment
of a magnon implies that it acquires a phase when propagating through an electric field.
This constitutes the Aharonov--Casher effect \cite{aharo84} analogous to the commonly known
Aharonov--Bohm effect for charges in a magnetic field \cite{aharo59}. Hence, magnonic Landau
levels can appear \cite{nakat17b, nakat17}. The electric field generated by a laser implies 
a time-periodic perturbation amenable to Floquet theory. This route to control
the effective Hamiltonian has recently been studied 
\cite{owerr17b, owerr18, nakat19, owerr19, owerr19b}.

The field of topological magnons has so far been strongly driven by transferring and adapting
ideas from fermionic systems. But one has to keep in mind essential persisting differences.
Fermionic bands can be completely filled or empty at zero temperature which is the basis of
the quantization of the quantum Hall conductivity \cite{kohmo85, avron83, niu85, uhrig91}.  
In bosonic systems, the notion of filled
bands does not exist, hence no quantization of the transversal conductivities is to be expected.
Furthermore, many electronic systems with topological properties preserve time-reversal symmetric systems 
so that the $\mathbb{Z}_2$ topological invariant applies. Magnetically ordered 
 systems, however, break time-reversal symmetry
so that the  definition of a $\mathbb{Z}_2$ topological invariant is only possible in combination
with additional symmetries. In this way, phenomena similar to the quantum spin Hall state can be 
established for magnetic excitations \cite{nakat17, lee18, kondo19, kondo19b, kawan19}.

\subsection{Triplon excitations}

Besides conventional long-range ordered system there exist many quantum antiferromagnets
which are valence bond solids. Certain bonds, dimer bonds, 
dominate over the other bonds.
For instance, the rung bonds in a spin ladder dominate over
the leg bonds \cite{shelt96a,schmi05b,krull12}. Then the elementary excitations
are triplons \cite{schmi03c}, i.e., dressed excitations of the dimer bonds from the singlet $S=0$ to the triplet state with $S=1$. In isotropic models, of course, the triplons are three-fold degenerate.

Strong frustration favors valence bond solids over long-range magnetic order
as in the Shastry--Sutherland lattice \cite{shast81b},
and hence favors triplons as elementary magnetic excitations. Their
mobility as expressed by their dispersion is strongly suppressed by
frustration, again impressively illustrated in the Shastry--Sutherland 
lattice \cite{miyah03,knett04a}. The reduced dispersion renders the degenerate
triplon bands prone to topological twists induced by small
perturbations such as DM terms inducing non-trivial topology 
\cite{romha15, malki17a, mccla17}. The dimerized honeycomb lattice 
can display non-trivial triplon bands \cite{anisi19} in form of a bilayer of two honeycomb lattices
which makes a triplon Hall effect possible \cite{joshi19}.

Topological behavior of triplons in 1D systems such as 
spin ladders has been predicted by investigating the Zak phase \cite{malki19c, nawa19} or the winding number \cite{joshi17}. In BiCu$_2$PO$_6$ and 
Ba$_2$CuSi$_2$O$_6$Cl$_2$ it has been verified by inelastic neutron scattering data \cite{malki19c,nawa19}. Topological triplons in chains may imply
localized end states, but only if they are protected by indirect
band gaps  \cite{malki19c}.

By extension of spin dimers with spin-$1$ triplons, certain lattices may also  
trimerize with spin-$1/2$ doublets and spin-$3/2$ quartets excitations 
which become topologically non-trivial if
a sufficiently strong DM term is present  \cite{romha19}.

\subsection{Spinon excitations}

The term `spinon' is used for fractional magnetic excitations of 
spin-$1/2$. Spinons appear also in correlated electronic systems where
spin and charge separate in a spinon and a holon
excitation. Hence, a number of studies addresses the question of
topological properties of Mott insulators, i.e., electronic systems
with sufficiently strong repulsive interactions so that they
turn insulating \cite{young08, pesin10, rache10, witcz10, karga11, cho12b,
ruegg12, yoshi14, scheu15, wang16c}. These approaches are constructed
to address systems without long-range magnetic order and without
spontaneous or non-spontaneous dimerization. A wide-spread caveat
consists in the use of mean-field approaches which start from 
the separation of spin and charge degrees of freedom. If one
follows this line of reasoning effective magnetic fields
engender Lorentz forces, Landau levels, and concomitant transversal
conductivities \cite{katsu10a, gao19}. 
Far from any isotropic spin models, for instance for
variants of the Kitaev model without any long-range order, deconfined 
spinon descriptions with topological properties can be established
 \cite{schaf15, sonne17}. 

\subsection{Skyrmion excitations}

Talking about topological properties in magnets, skyrmions have to 
be mentioned. However, the so far considered topological excitations
are itinerant and can be described by dispersive bands
of which the topology is captured by Berry phases. In contrast, skyrmions 
are static, long-lived magnetic textures in the nanometer range in
 FMs or AFMs which are topologically
protected. This means that their creation or annihilation requires
to overcome a substantial energy barrier. They are not
characterized by quantum Berry phases in momentum space, but by winding
numbers in real space such as 
\begin{align}
n = \frac{1}{4 \pi} \iint \mathrm{d} x \mathrm{d} y \, \vec{m} \cdot 
(\partial_x \vec{m} \times \partial_y \vec{m}) ,
\end{align}
where $m$ corresponds to the local magnetization. 
We do not go into detail because many excellent reviews of this
field exist \cite{nagao13, wiese16, kang16, finoc16, jiang17, fert17, evers18}. 

While skyrmions are {\it eo ipso} static, they can be moved
by various means, for instance by applying fields and/or currents.
In this way, they can store information, transport it, and may
 allow for processing it. This triggers abundant studies on 
applications, see above cited reviews.

Skyrmion lattices, i.e., periodic arrays of skyrmions, can be seen
to generate local magnetic field twisting the local quantization 
axes. This in turn generates Berry phases for the magnons,
the itinerant bosonic excitations on top of the skyrmion lattices,
for FM \cite{rolda16,garst17} and AFM \smash{systems \cite{diaz19, diaz20}.}
They show phenomena like the topological Hall effect \cite{neuba09, iwasa14}. 
This findings were triggered by the observation that a magnon can 
acquire a Berry phase by propagating along a closed contour in 
non-trivial magnetization profiles \cite{dugae05}.

\section{Experimental techniques and results}

While there are abundant theoretical studies on a large variety of
quantum magnets with topological properties, experimental observations 
are still scarce. So far, there are essentially two kinds
of studies, namely inelastic neutron scattering and transport measurements.
The dispersion of the itinerant magnetic excitations
is measured by inelastic neutron scattering (INS). This only provides
the eigenenergies, but no direct information on the eigenstates.
However, it is the momentum dependence of eigenstates which defines the
band topology, cf.\ Eq.\ \eqref{eq:berry-curva}. Hence,
only the comparison of the INS data to the results of 
theoretical models allows one to establish non-trivial topology.
In this way, topological magnons are confirmed in magnonic
insulators \cite{chisn15, chisn16} and in magnonic semimetal 
\cite{chen18, yao18, bao18, yuan20, zhang20}. Topological triplons 
are established in the Chern insulator SrCu$_2$(BO$_3$)$_2$ \cite{mccla17} and 
in quasi-1D spin ladders by their Zak phase \cite{malki19c, nawa19}.

Detecting transversal conductivities is the second experimental pillar 
to find non-trivial topology. Since magnetic excitations are neutral,
one aims at the thermal Hall effect. 
Separating phononic from magnonic effects is an issue in this respect.
If the temperature is low enough, the magnetic contribution dominates 
over the phononic one. So far, the thermal magnon Hall effect has been 
observed only for few material 
\cite{onose10, ideue12, hirsc15, tanab16, hentr19}. In a quantum spin ice magnet 
a thermal Hall effect was found and assigned to neutral excitations,
but they could not unambiguously identified as magnons or phonons 
\cite{hirsc15b}. For SrCu$_2$(BO$_3$)$_2$, no transversal conductivites
could be measured \cite{cairn20} in spite of the INS results \cite{mccla17}.

\section{Outlook and Conclusion}

This outlook represents a subjective view on ongoing or imminent
developments as well as important issues that need to be addressed in the authors' opinion.
Glancing at the above survey, it suggests itself that further experimental
progress is more urgent than theoretical one. Nevertheless, there are
also important theoretical challenges.

\subsection{Experimental Challenges}

It is desirable to establish a set of well-characterized model
compounds which allow for detailed experimental studies. 
With future applications in mind, the robustness and transport
properties of the edge states are important.
Efficient spin channels should display low dissipation, stability against disorder and be subjected to low Joule heating \cite{ruckr18}.
Up to now, more FM systems than AFM systems are known
because the generation and in particular the detection of magnetic
excitations are easier for the former. Yet, it is certainly indicated
to enforce the search for more AFM compounds.
Their linear dispersion at low energies implies much faster
propagation of magnetic excitations than in their FM counterparts.
In addition, the linear dispersion implies a significantly reduced
phase space for magnon decay into two or three magnons.
In FMs, the quadratic dispersion implies a continuum of scattering states
of two magnons into which a single magnon can decay.

Having information processing in mind, magnetic transport is the key
feature. Thus the energetically low-lying
excitations matter more than the high-lying ones unless 
high-lying excitations can be specifically generated.
So the specific generation and detection
of magnetic excitations is crucial. So far, bulk properties were the focus
of the investigations. But the most interesting physics goes 
on at the boundaries of topologically non-trivial phases. Hence, 
one needs experiments which can directly address the edge modes 
so that the signal does not need to be separated from a bulk 
signal. Clearly, this requires spatial resolution so that the 
boundaries can be excited separately. Furthermore, temporal 
resolution is required to measure velocities of signal 
transmission along the edges which constitutes a key quantity of 
edge modes \cite{uhrig16, malki17b, malki17c}. Then, control and manipulation of topological 
magnonic systems can be envisaged which will trigger a new boost 
to the field. A conceivable way to achieve spatial and temporal control is optical pumping by
inverse Faraday \smash{effect \cite{jackl17, savoc17}} or by parametric pumping
\cite{malz19}.

Finally, one wants to have interfaces between topological magnonics
and conventional electronics. To this end, spin-to-charge
conversion is a necessity for which first experimental 
results are available \cite{fiebi16, bossi18} and further theoretical
ideas are proposed \cite{zhang19b}.

\begin{figure}
\centering
\includegraphics[width=\columnwidth]{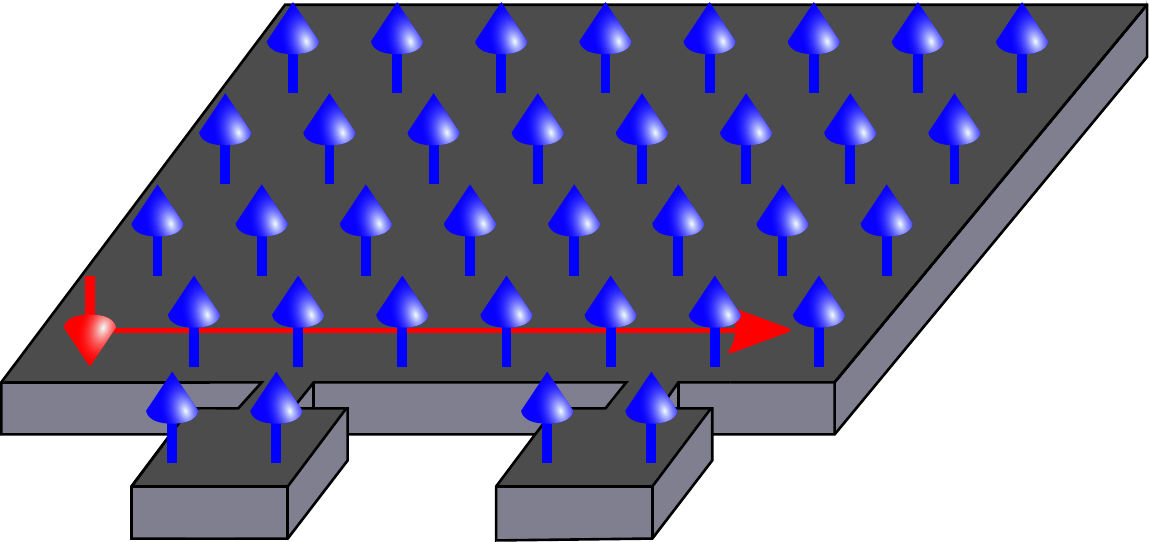} 
\caption{Schematics of a magnetic sample designed to reduce the signal transmission velocity in a controllable way. The shown spins indicate the ground state orientations and one magnon, not the lattice structure.}
\label{fig:mag_vel}
\end{figure}

Between generation and detection, the control of the magnonic 
information is essential. This implies switching signals, 
amplifying and delaying them, having them interfere with one another.
These issues are still fully at their infancy experimentaly. Yet, there exist ideas as to how spin-wave diodes, spin-wave beam splitters 
and spin-wave interferometers can be realized and 
controlled by manipulation of domain walls \cite{wang18c}.

Exemplarily, we outline a way to tune the velocity of signal transmission;
for magnetic systems this is the spin wave velocity. Tailored boundaries of
magnetic samples, see Fig.\ \ref{fig:mag_vel}, can reduce
this velocity by hybridizing propagating edge modes with local, non-dispersive
modes in the attached bays. For electronic systems 
this has been studied thoroughly for topological systems with chiral
edge modes 
\cite{uhrig16,malki17b, malki17c}. 
Adjusting the single-ion anisotropy \cite{panta18} in order to change the dispersion is a promising route.

\subsection{Theoretical Challenges}

The everlasting task of theory is to propose models
and mechanisms which explain experimental observations. In addition,
theory can propose setups and systems to get better handles
on the relevant experimental signatures. Furthermore, there are also
intrinsic theoretical issues.

Conceptually, damping or decay of the excitations which
carry the information is a crucial issue. The signal damping 
is an experimental fact so that any measurement needs to be faster than the decay of the signal. But there resides
also a deeper conceptual issue: how is a Berry phase properly
defined if there are no well-defined elementary particles
because they hybridize with the continua of scattering states? 
The first steps are undertaken 
\cite{chern16,persh18,mcclar18,mcclar19}, but deeper insight is still missing.
Strong magnon-magnon or triplon-triplon interactions
can give rise to bound states. Their interplay with the
elementary magnons or triplons also needs to be taken into
account \cite{qin17c, qin18b}.

In addition, the interaction between magnetic and phononic degrees of freedom
can also play an important role either by 
leading to topological magnons \cite{zhang19} 
or to topological phonons \cite{thing19} or to topological \smash{polarons \cite{park19}.}

\subsection{Conclusion}

On the one hand, topological magnonics, i.e., using dispersive
magnetic excitations for information storage, transmission
and processing, is a very promising and fundamentally exciting
research area. On the other hand, the experimental and theoretical issues
to be tackled and solved are challenging. This should trigger
our curiosity and ingenuity to come up with seminal solutions!

\paragraph{Acknowledgments} 
This work was supported by the Deutsche Forschungsgemeinschaft in project 
UH 90/14-1.


%

\clearpage
\pagebreak

\setcounter{equation}{0}
\setcounter{figure}{0}
\setcounter{table}{0}
\setcounter{page}{1}
\makeatletter
\renewcommand{\theequation}{S\arabic{equation}}
\renewcommand{\thefigure}{S\arabic{figure}}
\renewcommand{\bibnumfmt}[1]{[S#1]}
\renewcommand{\citenumfont}[1]{S#1}

\vspace*{0.5cm}

\onecolumngrid
\begin{center}
\textbf{\large  Supplemental Material for "Topological Magnetic Excitations"}
\end{center}

This supplement consists of four sections. (I) We provide the symmetry analysis of spin lattices according to the five selection rules of Moriya. (II) We illustrate the derivation of the Dzyaloshinskii-Moriya interaction from spin-orbit coupling.  (III) A ferromagnetic Heisenberg model on a honeycomb lattice illustrates the similarity between electronic and magnonic systems. 
By applying a bosonic spin representation we deduce a magnon model analogous to the Haldane model.
(IV) We introduce the concept of the Berry phase  and illustrate it by the example of the Su-Schrieffer–Heeger model. The latter model also enables us to illustrate boundary states
as implied by the bulk-boundary correspondence.

\vspace{1cm}
\twocolumngrid

\section{Dzyaloshinskii-Moriya Interaction} 
\label{supp:one}

The Dzyaloshinskii-Moriya (DM) interaction can be seen as the main
driving mechanism to introduce a topological twist to eigenstates of magnetic systems. 
The point group symmetries of the model restrict the directions of the DM vectors 
$\vec{D}$ 
which in turn  determine the DM interactions according to the five selection rules
formulated by \smash{Moriya \cite{moriy60b}.} Here, we briefly present these five selection rules
considering two coupled ions with spin. One spin is at site $A$  interacting with the 
spin at site $B$. The center point of the connecting line $\overline{AB}$ is labeled by $C$.

\begin{enumerate}
\item If $C$ is a center of inversion, $\vec{D} = 0$ is implied. 
\item In case a mirror plane perpendicular to $\overline{AB}$ exists and comprises $C$,
 $\vec{D} \perp \overline{AB}$ holds so that $\vec{D}$ lies in this mirror plane.
\item If a mirror plane comprises sites $A$ and $B$, the vector $\vec{D}$ 
must be perpendicular to it.
\item If a twofold rotation axis exists perpendicular to $\overline{AB}$ and passing through $C$,  the vector $\vec{D}$ is parallel to this rotation axis. 
\item In case of an $n$-fold rotation axis ($n \geq$ 2) along $\overline{AB}$
exists, $\vec{D} \parallel \overline{AB}$ holds. 
\end{enumerate}

As stated the selection rules are based on inversion, mirror and rotation symmetry. 
The degree of restriction of the DM vector increases for increasing symmetry. 
Thus, lower symmetry is favorable for topological effects. 

\section{Derivation of the Dzyaloshinskii-Moriya Interaction}
\label{supp:two}

The DM interaction results from Anderson superexchange if a spin-orbit coupling, i.e., 
a spin dependent hopping of electrons, is present  
\cite{shekh92, shekh93, moriy60a, moriy60b}. The calculation below
is included to emphasize the connection to spin dependent hopping, but 
keeping it as simple as possible. The aim is to derive the term
\begin{align}
\mathcal{H}_\mathrm{DM} &= \sum_{\langle i, j\rangle} 
\vec{D}_{ij} \cdot (\vec{S}_i \times \vec{S}_j)  
\label{eq:dm_ham}
\end{align}
for two sites $i\ne j$ where we restrict ourselves to nearest-neighbour sites 
$\langle i, j\rangle$ for clarity.
The fermionic Hamiltonian belongs to a Hubbard model and is 
split into a local on-site Hamiltonian $\mathcal{H}_0$ and a hopping Hamiltonian $\mathcal{H}_1$
{ \setlength{\mathindent}{14pt}
\begin{subequations}
\begin{align}
\mathcal{H}_0 &= E \sum_{{i,\sigma}} a^\dagger_{i,\sigma}  
a^{\phantom{\dagger}}_{i,\sigma}
 + U \sum_{i} a^\dagger_{i,\uparrow}  
a^{\phantom{\dagger}}_{i,\uparrow} a^\dagger_{i,\downarrow} 
a^{\phantom{\dagger}}_{i,\downarrow} 
\\
\mathcal{H}_1 &= t \sum_{\langle i, j\rangle,\sigma}  a^\dagger_{i,\sigma}  
a^{\phantom{\dagger}}_{j,\sigma} + \sum_{\langle i, j\rangle,\alpha\beta} 
 \vec C_{ij} \cdot \vec \sigma_{\alpha\beta}\;  a^\dagger_{i,\alpha} 
a^{\phantom{\dagger}}_{j,\beta} \phantom{a} ,
\end{align}
\end{subequations}}where $E$ is the local single-particle energy, $t$ the isotropic hopping element, $U$ the local
Hubbard repulsion, $\vec{\sigma}$ is the usual vector of
Pauli matrices, $\alpha,\beta \in \{\uparrow,\downarrow\}$ and the \smash{vector $\vec C_{ij}$} determines
the spin dependent hopping resulting from spin orbit coupling.

The main step is to apply perturbation theory to the half-filled states
with precisely one electron at each site in second order of $\mathcal H_1$. The 
resulting effective Hamiltonian is given by
\begin{align}
\mathcal{H}_\mathrm{eff} = P_0 \mathcal{H}_1 P_1(E_0 - \mathcal{H}_0)^{-1} P_1\mathcal{H}_1 P_0
\end{align} 
where $P_n$ is the projection operator onto the Hilbert subspace with precisely $n$ double occupancies and $E_0$ the ground state energy. 
The hopping term automatically creates one double occupancy so that 
$(E - \mathcal{H}_0)^{-1} = (-U)^{-1}$ holds. 
The second application of $\mathcal{H}_1$ annihilates the double occupancy so that
the system reaches again the states with all sites singly occupied. 
States with additionally created double occupancies are projected out by $P_0$.
If both hoppings
are isotropic, the known antiferromagnetic superexchange $J=4t^2/U$ results
\begin{equation}
H_\text{isotropic} = J\sum_{\langle i, j\rangle} \vec S_i\cdot\vec S_j \quad ,
\end{equation}
where we use
\begin{subequations}
\begin{align}
a^\dagger_{i,\uparrow}  a^{\phantom{\dagger}}_{i,\uparrow} - 
a^\dagger_{i,\downarrow}  a^{\phantom{\dagger}}_{i,\downarrow} &= 2 S^z_i 
\\
a^\dagger_{i,\uparrow} a^{\phantom{\dagger}}_{i,\downarrow} &= S^+_i
\\
a^\dagger_{i,\downarrow} a^{\phantom{\dagger}}_{i,\uparrow} &= S^-_i \quad .
\end{align}
\end{subequations}
 But in presence of spin-dependent hopping one hopping can be isotropic ($t$) 
and the other spin-dependent ($\vec C$) leading to the antisymmetric 
DM coupling in Eq.~\eqref{eq:dm_ham} with
\begin{align}
\vec{D}_{ij} &=  \frac{4 \mathrm{i}}{U} \left[t\vec{C}_{ij} - t\vec{C}_{ji}  \right] \quad . 
\end{align}
There are also symmetric couplings arising in second order in $\vec C$, but we
do not discuss them here and refer 
the interested reader to specialized references \cite{shekh92, shekh93}.

\section{Similarity between Electronic and Magnonic Models}
\label{supp:three}

In order to clearly see the similarity between electronic and magnonic systems it is helpful
 to express the spin operators  in terms of elementary excitations of bosonic nature. 
For illustration, we present the example of topological magnons in a ferromagnetic honeycomb lattice 
\cite{owerre16c, owerr16}. The aim is to show that single magnons in this model essentially behave like single
 fermions in the Haldane model. We choose the Haldane model \cite{halda88b} because it is one
of the first and paradigmatic fermionic models with non-trivial topology and often used for proof-of-principle
studies.

The spin Hamiltonian consists of three parts
\begin{subequations}
\begin{align}
\mathcal{H} &= \mathcal{H}_{\mathrm{NN}} + \mathcal{H}_{\mathrm{NNN}} + \mathcal{H}_{\mathrm{DM}}
\\
\mathcal{H}_{\mathrm{NN}}  &= - J \sum_{\langle  ij \rangle} \vec{S}_i \cdot \vec{S}_j 
\\
\mathcal{H}_{\mathrm{NNN}}  &=
- J' \sum_{\langle \langle ij \rangle\rangle}\vec{S}_i \cdot \vec{S}_j 
\\ 
\mathcal{H}_{\mathrm{DM}}  &= 
\sum_{\langle\langle ij \rangle\rangle} \vec{D}_{ij} \cdot  (\vec{S}_i \times \vec{S}_j  )
\end{align}
\end{subequations}
with two isotropic Heisenberg couplings $J$ and $J'$ and one DM coupling $\vec{D}_{ij} $. 
The isotropic couplings are ferromagnetic $J, J' > 0$ while pairs of nearest neighbors (NN) and next-nearest neighbors (NNN) are denoted by 
$ \langle ij \rangle$ and by $\langle \langle ij \rangle\rangle$, respectively. For clarity, the DM couplings are restricted to a direction perpendicular 
to the plane of the lattice $\vec{D}_{ij} = \nu_{ij} D_z \hat{e}_z$ with $\nu_{ij} = \pm 1$. 
The anti-clockwise hopping from site $i$ to site $j$ around a honeycomb is counted positive while the clockwise hopping is counted negative. 
We assume a fully polarized ground state in $z$ direction even in presence of DM interaction. 

There are several bosonic representations of the spin operators. In leading order in $1/(2S)$, both the Dyson--Maleev representation 
\cite{dyson56a, malee57} and the Holstein-Primakoff respresentation \cite{holst40} correspond to
\begin{subequations}
\begin{align}
S^+_i &= \sqrt{2S} ( b_i + \mathcal O((2S)^{-2})\\
S^-_i &= \sqrt{2S} (b_i^\dagger  + \mathcal O((2S)^{-2}) \\
S^z &= b_i^\dagger b_i^{\phantom{\dagger}} - S \quad .
\end{align}
\end{subequations}
In order to stay in a one-particle framework and to focus on low-lying excitations 
we use the harmonic approximation which means that
we keep only bilinear terms. Thus, the first Heisenberg term is expressed as 
{ \setlength{\mathindent}{6pt}
\begin{subequations}
\begin{align}
&\mathcal{H}_{\mathrm{NN}} = - J \sum_{\mathclap{\left\langle  ij  \right\rangle}} \left[ \frac{1}{2} (S_i^+ S_j^- + S_i^- S_j^+) + S_i^z S_j^z \right] 
\\
&\quad = - J S \sum_{\mathclap{\left\langle  ij  \right\rangle}} 
\left[ (b_i^\dagger b_j^{\phantom{\dagger}} + b_j^\dagger b_i^{\phantom{\dagger}}) 
- (b_i^\dagger b_i^{\phantom{\dagger}} + b_j^\dagger b_j^{\phantom{\dagger}} ) + S \right] \phantom{a} .
\label{eq:heisen1}
\end{align}
\end{subequations}} The first round bracket in Eq.\ \eqref{eq:heisen1} describes NN hopping of magnons 
on the honeycomb lattice while the second round bracket
shifts the dispersion. The NNN Heisenberg couplings yield the analogous result. The same procedure is applied to the DM interaction
\begin{subequations}
\begin{align}
\mathcal{H}_{\mathrm{DM}} =& \sum_{\mathclap{\left\langle \left\langle ij  \right\rangle \right\rangle}} 
\vec{D}_{ij} \vec{S}_i \times \vec{S}_j 
\\
 =& - \frac{\mathrm{i} D_z}{2} \sum_{\mathclap{\left\langle \left\langle ij  \right\rangle \right\rangle}} 
\nu_{ij} ( S_i^+ S_j^- - S_i^- S_j^+ ) 
\\
 =& - \mathrm{i} D_z S \sum_{\mathclap{\left\langle \left\langle ij  \right\rangle \right\rangle}} 
\nu_{ij} (  b_i^\dagger b_j^{\phantom{\dagger}} - b_j^\dagger b_i^{\phantom{\dagger}} ) \quad .
 \label{eq:dm1}
\end{align}
\end{subequations}
Combining both hopping terms to the NNN yields the complex hopping known from the Haldane model 
\begin{subequations}
\begin{align}
\left[\mathcal{H}_{\mathrm{NNN}} + \mathcal{H}_{\mathrm{DM}}\right]_\text{hop}  &=
- S \sum_{\mathclap{\left\langle \left\langle ij  \right\rangle \right\rangle}} 
(J' + \nu_{ij}\mathrm{i} D_z) b_i^\dagger b_j^{\phantom{\dagger}} 
\\
&\hspace{-5mm}= -S \sqrt{J'^2 + D_z^2}  \sum _{\mathclap{\left\langle \left\langle ij  \right\rangle \right\rangle}} 
\mathrm{e}^{\mathrm{i} \nu_{ij} \phi} b_i^\dagger b_j^{\phantom{\dagger}}
\end{align}
\end{subequations}
with $\phi = \mathrm{arctan} (J'/D_z)$. This shows how a finite DM interaction induces a complex hopping for magnons
on the honeycomb lattice which is responsible for the topological twist. Eventually, we obtain the Haldane model in terms of 
bilinear bosonic operators representing the magnons of the original spin Hamiltonian. 
The constant on-site terms do not affect the topology in the bulk because they do not influence
the eigenstates. Similarly, combinations of isotropic Heisenberg and anisotropic DM interactions 
can be used on any lattice model to emulate electronic models by a spin model on the level of single elementary excitations. 
In analogy, the crossproduct  in the scalar spin chirality can act like the DM interaction. 

In case of antiferromagnetic couplings on bipartite lattices, similar calculations can be done once one has
mapped the N\'eel state to the fully polarized state by rotating the spins of one sublattice by 180$^\circ$. 
But even on the  bilinear level, terms creating or annihilating pairs of bosons appear which break the conservation of
particle number. Still, a suitable Bogoliubov transformation diagonalizes the model yielding well-defined dispersions
and eigenstates. The necessary Bogoliubov transformations are no longer represented by unitary matrices so that 
the subsequent calculation of Berry phases and Chern numbers requires to use a generalization of the standard scalar product 
to a so-called `symplectic' or `paraunitary' product for the involved operators, 
see for instance the Appendix of Ref.\
 \cite{malki19c} for a comprehensive presentation of this subtlety.

Of course, the harmonic, bilinear approximation used above neglects effects such as two-particle interactions and spontaneous
quasi-particle decay. Furthermore, one still has to keep in mind that the bosonic quasi-particles obey another statistics
than fermions. Nevertheless, one expects that  particular topological properties of the magnetic excitations are retained.

Finally, we point out that besides the representation of elementary excitations in ordered spin systems by 
ferromagnetic or antiferromagnetic magnons the elementary excitations in disordered spin systems can often be 
described by triplon operators, see main text. For them, DM interactions lead similarly
 to complex hopping processes yielding topological twists.

\section{Berry Phase and Boundary States}
\label{supp:four}

The Berry phase \cite{berry84} is one of the fundamental concepts to describe topological characteristics. For instance the quantum Hall effect is classified by topological \smash{invariants 
\cite{kohmo85, avron83, niu85, uhrig91}} based on the Berry phase. We will introduce the notion of the Berry phase and briefly present 
the example of the Su-Schrieffer–Heeger (SSH) model \cite{Su80, heege88} 
to gain deeper insight. 

We start by considering an adiabatic evolution of a general Hamiltonian depending on a parameter set $\vec{P}$ along a closed path $\Gamma$. According to the adiabatic theorem the non-degenerate eigenstate remains in its instantaneous eigenstate if the variation of continuous parameters is slow enough and the considered eigenstate is
separated from other states by an energy gap. The instantaneous eigenstates of the Hamiltonian can be calculated by diagonalization of 
\begin{align}
H(\vec{P}) \ket{n, \vec{P}} = E_n (\vec{P}) \ket{n, \vec{P}} \quad ,
\end{align}
at each point $\vec{P}$ in
parameter space where $n$ is the label of the energy eigenstates. Note that the instantaneous eigenstates are not completely determined by this equation
 since nothing fixes their phase. In order to remove this arbitrariness we assume that
 a smooth gauge of the phase can be found such that 
$f(\vec P,\vec P'):= \langle n, \vec{P} \ket{n, \vec{P'}}$ is a differentiable function of both arguments. The Berry phase $\gamma_n$ depends only on the contour $\Gamma$ in parameter space along which the system is changed and on the state $n$ considered. It also depends on the phase gauge unless the contour $\Gamma$ is closed.

To see how the Berry phase appears we follow the Gedankenexperiment that the
system is varied in time in parameter space along the closed path $\vec{P}(t)$
starting at $t=0$ and finishing at $t=t_f$ with $\vec{P}(t_f)=P(t=0)$.
According to the adiabatic theorem one can 
 describe the time evolution of the state by the ansatz
\begin{align}
\ket{\psi(t)} = \exp(\mathrm{i} \theta(t)) \ket{n, \vec{P}(t)} 
\end{align}
with a smooth phase $\theta(t)$. Inserting the ansatz into the 
the Schrödinger equation ($\hbar = 1$) and multiplying with $\bra{n, \vec{P}(t)}$
from the left yields
\begin{align}
- E_n(\vec{P}(t)) + \mathrm{i} \bra{n, \vec{P}(t)} \frac{\mathrm{d}}{\mathrm{d} t} 
\ket{n, \vec{P}(t)} 
	= \frac{\mathrm{d}}{\mathrm{d} t} \theta(t) 
\end{align}
which is solved by 
\begin{subequations}
\begin{align}
\label{eq:dyn-phase}
\theta(t) &= - \int_0^t E_n(\vec{P}(t')) \mathrm{d} t' 
\\
\label{eq:geo-phase}
& + \mathrm{i} \int_0^t \bra{n, \vec{P}(t')}
	\frac{\mathrm{d}}{\mathrm{d} t'} \ket{n, \vec{P}(t')} \mathrm{d} t' \quad .
\end{align}
\end{subequations}
The first term \eqref{eq:dyn-phase} is the dynamic phase growing linearly in
time if the process is extended to longer and longer times, i.e., 
performed slower and slower. The second term \eqref{eq:geo-phase} describes
the geometric phase; this is the Berry phase. Note that it does \emph{not} matter
how fast the path $\vec P(t)$ is followed. 
The ostensible time dependence is removed if the explicit dependence on $\vec{P}$ is used instead 
\begin{equation}
\frac{\mathrm{d}}{\mathrm{d} t'} \ket{n, \vec{P}(t')} \mathrm{d} t'
= \vec\nabla_{\vec P} \ket{n, \vec{P}}
\end{equation}
yielding eventually
\begin{subequations}
\begin{align}
\gamma_n &=  \mathrm{i} \int_\Gamma \bra{n, \vec{P}}  
\vec\nabla_{\vec{P}} \ket{n, \vec{P}} \cdot \mathrm{d} \vec P
\\
 &= \int_\Gamma \vec A_n(\vec{P})\cdot \mathrm{d} \vec{P} \label{eq:an}
\\
\vec A_n(\vec{P}) &:=  \mathrm{i} \bra{n, \vec{P}}  
\vec\nabla_{\vec{P}} \ket{n, \vec{P}}\quad ,
\end{align}
\end{subequations}
where the quantity $\vec A_n(\vec{P})$ is called Berry connection.

\begin{figure}
\centering
\includegraphics[width=\columnwidth]{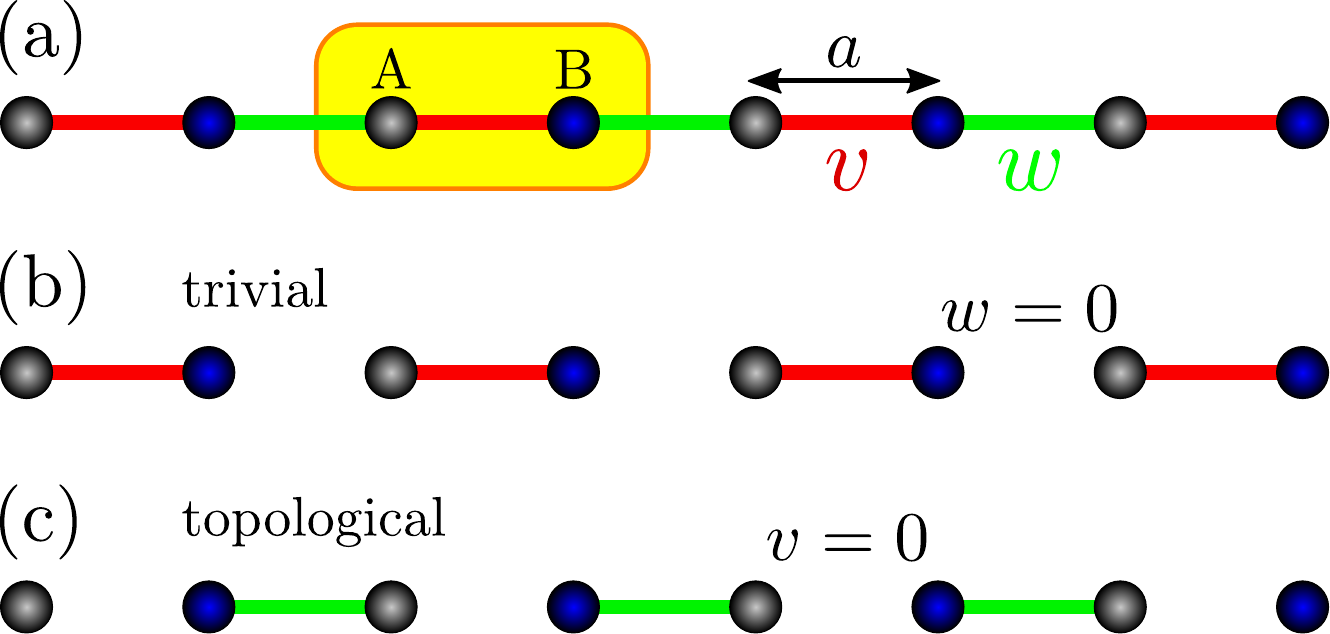} 
\caption{(a) Schematic sketch of the chain of the Su-Schrieffer-Heeger model. 
The intracell and intercell hoppings are denoted by $v$ and $w$, respectively. 
The unit cell is highlighted by  yellow shading. 
The two limiting cases are shown in (b) and (c).}
\label{fig:ssh}
\end{figure}

If one considers the closed contour around the Brillouin zone, i.e., $\vec P =\vec k$,
 one can define a topological invariant from the Berry phase. For instance, we consider the 
one-dimensional Su-Schrieffer-Heeger (SSH) model which is described by the
 tight-binding model
\begin{align}
\label{eq:ssh}
	\mathcal{H}_\mathrm{SSH} &= \sum_i \left( v c_{i, \mathrm{B}}^\dagger 
	c_{i, \mathrm{A}}^{\phantom{\dagger}} + 
	w c_{i+1, \mathrm{A}}^\dagger c_{i, \mathrm{B}}^{\phantom{\dagger}} \right) 
	+ \mathrm{h.c.} \quad .
\end{align}
The bipartite chain is shown in Fig.\ \ref{fig:ssh} where the intracell hoppings 
$v$ and intercell hoppings $w$ are displayed by red and green bonds, respectively. 
The unit cell is shown by the yellow shaded box and 
the lattice constant is given \smash{by $a$.} 
The $2\times 2$ matrix representation $H_k$ of the Hamiltonian at given wave vector $k$ can be denoted by
\begin{equation}
H_k = \vec d\cdot\vec\sigma \quad, 
\end{equation}
where $\vec\sigma$ is the vector of Pauli matrices and
\begin{align}
\vec{d}(k) := (v + w \cos(k), w \sin(k), 0)
\end{align}
resulting from Fourier transformation of \eqref{eq:ssh}.
The bulk dispersion of the two bands is easily found  
\begin{align}
E_\pm(k) &= \pm d = \pm \sqrt{d_x^2+d_y^2+d_z^2} \quad .
\label{eq:eigenenergies}
\end{align}
The two bands are separated for  $v \neq w$ while $v=w$ defines 
the transition line between the two topological phases of the SSH model
where the energy gap between both bands closes.
The two corresponding eigenstates can be expressed by
\begin{align}
	\ket{\pm} &= \frac{1}{\sqrt{2d(d\pm d_z)}} 
	\begin{pmatrix}
		d_z \pm d \\
		d_x - \mathrm{i} d_y	
	\end{pmatrix} \quad .
\end{align}
Consequently, the Berry connection is  given by 
\begin{subequations}
\begin{align}
	A_{\pm} &= \mathrm{i} \bra{\pm} \partial_{k} \ket{\pm} 
	\\
	&= \frac{-1}{2d(d \pm d_z)} (d_y\partial_{k} d_x - d_x\partial_{k} d_y) \quad .
\end{align}
\end{subequations}
The Berry phase is calculated by integration of $A_\pm$ over $k$ from $0$ to $2 \pi$ as defined in 
Eq.\ \eqref{eq:an}. For $|v| > |w|$ the Berry phase is $0$ while for 
$|v| < |w|$ it takes the finite value $\pm\pi$ which corresponds to a nontrivial topological character
of this quantum phase. Note that swapping \smash{$v\leftrightarrow w$} does not change the bulk dispersion,
but does change the eigenvectors. This exemplifies that the topological twist is encoded in the eigenstates of the system and not in its eigenenergies.

The SSH model also allows us to illustrate the bulk-boundary correspondence by
considering finite chains in the two limiting cases shown in Fig.\ \ref{fig:ssh}(b) and (c).
The non-trivial topological case for $v = 0$ displays localized edge states at the ends of the 
finite piece of chain while the topologically trivial case $w = 0$  leads only to isolated dimers.
This illustrates that generically the quantum phase with non-trivial Berry phase $\gamma_n\ne 0$
displays localized boundary states while the trivial phase with Berry phase $\gamma_n= 0$
does not. But we emphasize that the topological edge states may delocalize
in spite of non-trivial topological invariants if they are not protected additionally 
by an indirect energy
gap, for details see Ref.\ \cite{malki19c,malki19b}.


%

\end{document}